\documentclass[twocolumn,showpacs,preprintnumbers,nofootinbib,amssymb,10pt]{revtex4-1}

\usepackage{graphicx, epsfig, amssymb} 
\usepackage{amsmath, amsfonts}
\usepackage{bm} 

\usepackage[linktocpage]{hyperref}
\usepackage[caption=false]{subfig}
\usepackage[usenames]{color}

\def\be{\begin{equation}}
\def\ee{\end{equation}}

\newcommand{\bea}{\begin{eqnarray}}
\newcommand{\eea}{\end{eqnarray}}
\newcommand{\ben}{\begin{enumerate}}
\newcommand{\een}{\end{enumerate}}
\newcommand{\bi}{\begin{itemize}}
\newcommand{\ei}{\end{itemize}}

\def\ga{\mathrel{\raise.3ex\hbox{$>$\kern-.75em\lower1ex\hbox{$\sim$}}}}
\def\la{\mathrel{\raise.3ex\hbox{$<$\kern-.75em\lower1ex\hbox{$\sim$}}}}

\def\l{\left}
\def\r{\right}
\def\be{\begin{equation}}
\def\ee{\end{equation}}

\def\I_M{{I_{\scriptscriptstyle M\times M}}}

\def\be{\begin{equation}}
\def\ee{\end{equation}}
\def\bea{\begin{eqnarray}}
\def\eea{\end{eqnarray}}
\newcommand{\beq}{\begin{eqnarray}}
\newcommand{\eeq}{\end{eqnarray}}
\def\pa{\partial}

\newcommand{\beqal}{\begin{eqnarray}\label}
\newcommand{\beqa}{\begin{eqnarray}}
\newcommand{\eeqa}{\end{eqnarray}}

\begin{document}
\title{Absorption of planar massless scalar waves by Bardeen regular black holes}

\author{Caio F. B. Macedo}\email{caiomacedo@ufpa.br}
\author{Lu\'is C. B. Crispino}\email{crispino@ufpa.br}
\affiliation{Faculdade de F\'{\i}sica, Universidade 
Federal do Par\'a, 66075-110, Bel\'em, Par\'a, Brazil.}

\begin{abstract}
Accretion of fields by black holes is a subject of great interest in physics. 
It is known that accretion plays a fundamental role in active galactic nuclei and in the evolution of black holes. Accretion of fundamental fields is often related to the study of absorption cross section.
Basically all black holes for which absorption of fields has been studied so far present singularities.
However, even within general relativity, it is possible to construct regular black holes: objects with event horizons but without singularities. Many physically motivated regular black hole solutions have been proposed in the past years, demanding the understanding of their absorption properties.
We study the absorption of planar massless scalar waves by Bardeen regular black holes. 
We compare the absorption cross section of Bardeen and Reissner--Nordstr\"om black holes, showing that the former always have a bigger absorption cross section for fixed values of the field frequency and of the normalized black hole charge. 
We also show that it is possible for a Bardeen black hole to have the same high-frequency absorption cross section of a Reissner--Nordstr\"om black hole. Our results suggest that, in mid-to-high-frequency regimes, regular black holes can have compatible properties with black holes with singularities, as far as absorption is concerned.

\end{abstract}

\date{\today}

\maketitle

\section{Introduction}

One of the most intriguing predictions of general relativity (GR) is the existence of black holes (BHs). BHs became a paradigm in physics, and are believed to populate the galaxies \cite{Narayan:2005ie}. Within standard GR, black holes are simple objects, described only by their mass, angular momentum and charge \cite{heuslerreview}. However, standard black holes suffer from one of the main problems of GR: the presence of singularities. Our physical knowledge breaks down at singularities. Although generally hidden by a horizon, and protected by the Penrose conjecture \cite{Penrose:1969pc} (see also \cite{Wald:1997wa} for a review), singularities are expected to exist within GR, according to the singularity theorems developed by Hawking and collaborators \cite{hawkingb}.

Singularities are expected to be better understood with an improved theory of gravity (whether an extension or a modification of GR) \cite{Clifton:2011jh}. Notwithstanding, within GR it is possible to obtain BH solutions without singularities. Bardeen presented a BH solution without singularities that satisfies the weak energy condition in GR \cite{Bardeen}. Although Bardeen's solution has its theoretical motivation in the studies of BH spacetimes with no singularities, a stronger physical motivation for it was missing until it was shown that the Bardeen BH is a solution of GR with a nonlinear magnetic monopole, i.e., a solution of the Einstein's equations coupled to a nonlinear electrodynamics \cite{AyonBeato:2000zs}. Apart from this, further works with other physically motivated regular BHs can be found in the literature (see, e.g., Refs. \cite{Ansoldi:2008jw,Lemos:2011dq}).

One way to test the physics of BHs is analyzing test fields around them. In this context, there are the quasinormal modes: natural oscillation frequencies of the fields with physically motivated boundary conditions \cite{Nollert:1999ji,Berti:2009kk}. An extensive survey of quasinormal modes of test charged scalar fields around different types of regular BHs was presented in Ref. \cite{Flachi:2012nv}. Quasinormal modes of the Dirac field were investigated in Ref. \cite{Li:2013fka} and of the massive scalar fields in Hayward regular BHs in Ref. \cite{Lin:2013ofa}. Quasinormal modes have an interesting relation with scattering processes in BH spacetimes. This relation can be seen, for instance, in the scattering of Gaussian packets by BHs \cite{Vishveshwara:1970zz,Andersson:1996cm}.

Another important aspect of BHs is how they absorb matter and fields around them, i.e., their accretion rate. Accretion has an important role in the phenomenology of active galactic nuclei, and can be considered as an important agent to the mass growth of their BH hosts (see, e.g., Refs. \cite{Granato:2003ch,Marconi:2003tg,Ferrarese:2004qr} and the references therein).
Along more than 45 years, absorption of scalar fields has been studied extensively in many BH scenarios 
(see, e.g., Refs.
 \cite{Matzner:1968,Sanchez:1977si,Futterman:1988ni,Higuchi:2001si,Higuchi:2001sib,Jung:2005mr,Macedo:2013afa,Benone:2014qaa} and the references therein). 
The initial field configuration is usually taken to be plane waves at infinity and the problem is often directed to compute the absorption cross section of the field. Also, in the classical (high-frequency) limit, absorption cross sections are directly related with the shadows of BHs \cite{Huang:2007us,Johannsen:2010ru,Li:2013jra}. Moreover, the case of planar waves absorption has many features in common with the case of accretion of a fluid moving with constant velocity toward a BH (see, e.g., Ref. \cite{Petrich:1988zz}), which turns out to be important in the phenomenology of extreme mass-ratio inspirals \cite{Macedo:2013qea,Macedo:2013jja}.

In this paper we address the problem of how regular BHs absorb fields, focusing in the analysis of the absorption cross section of planar massless scalar waves by a Bardeen regular BH. Generically, the line element of spherically symmetric BH spacetimes can be written in the standard spherical coordinate system as
\be
ds^2=-f(r)dt^2+f(r)^{-1}dr^2+r^2d\Omega^2, \label{eq:line}
\ee	
where the function $f(r)$ depends on the particular BH under consideration. As we shall see, the Bardeen BH has a structure very similar to that of a standard electrically charged BH within GR, i.e., of a Reissner--Nordstr\"om (RN) BH. Because of that, we shall compare our results with the RN BH ones \cite{Jung:2005mr,PhysRevD.79.064022}.

The remainder of this paper is organized as follows. In Sec. \ref{sec:bardeenbh} we review some aspects of the Bardeen regular BHs. In Sec. \ref{sec:absorption} we revisit the main aspects of the absorption cross section of planar massless scalar waves in spherically symmetric BH spacetimes. We also present the results in the low- and high-frequency regimes for the massless scalar absorption cross section of Bardeen BHs. In Sec. \ref{sec:results} we exhibit a selection of our numerical results. We compare our results for the Bardeen regular BH with the results for the RN BH. Also, we discuss the possibility of having a Bardeen BH with a similar absorption cross section of a RN BH. We present our final remarks in Sec. \ref{sec:final}. Throughout the paper we use natural units, for which $G=c=\hbar=1$.

\section{Bardeen 
regular black holes}\label{sec:bardeenbh}
As mentioned in the Introduction, the Bardeen BH was one of the first regular BH solutions presented in the literature~\cite{Bardeen}. Later, it received the physical interpretation of a BH with a nonlinear magnetic monopole \cite{AyonBeato:2000zs}. Nonlinear electrodynamics theories within GR are generically described by the action
\be
S=\int d^4x\sqrt{-g}\l[\frac{1}{16 \pi}R-\frac{1}{4 \pi}\mathcal{L}(F)\r],
\label{eq:fieldaction}
\ee
 where $R$ is the Ricci scalar; $\mathcal{L}$ is the Lagrangian of the electromagnetic field; $F=\frac{1}{4}F_{ab}F^{ab}$; with $F_{ab}$ being the standard electromagnetic field strength; and $g$ is the determinant of the metric $g_{ab}$. For the theory that generates the Bardeen regular BH, we have that
\be
\mathcal{L}(F)=\frac{3}{2\,s\,q^2}\l(\frac{\sqrt{2 q^2 F}}{1-\sqrt{2 q^2 F}}\r)^{5/2},
\ee
where $s=|q|/(2 M)$, $q$ is the magnetic charge and $M$ is the mass of the configuration  \cite{AyonBeato:2000zs}. The line element of the Bardeen BH is given by Eq. \eqref{eq:line}, with
\be
f(r)=1-\frac{2M r^2}{(r^2+q^2)^{3/2}}.
\ee
 The Bardeen solution has a structure similar to the RN spacetime, presenting two horizons up to some value of the BH charge. For $q=q_{\text{ext}}=4M/(3\sqrt{3})$, the two horizons coincide and we have the so-called extremal BH. In this paper, we shall consider $0\leq q\leq q_{\text{ext}}$.	

For later comparison, it is instructive to mention explicitly the RN solution. The line element of the RN spacetime is given by Eq. \eqref{eq:line}, with
\be
f(r)=1-\frac{2M}{r}+\frac{q^2}{r^2},
\ee
where, in this case, $q$ is the electric charge of the BH. The extreme case of the RN BH is given by $q_\text{ext}=M$. Note that we are using the same symbol ($q$) for both magnetic (Bardeen BH) and electric (RN BH) charge. To better compare both spacetimes, we shall present our results in terms of the normalized charge ${Q}\equiv q/q_\text{ext}$.

\section{Absorption cross section} \label{sec:absorption}

\subsection{Partial-waves approach}

A massless scalar field $\Phi$ is described by the Klein--Gordon equation, namely,
\be
 \frac{1}{\sqrt{-g}}\pa_a \l(\sqrt{-g}g^{ab}\pa_b\Phi\r)=0.
\label{kleingordon}
\ee
Here we are considering a minimally coupled scalar field.

\begin{figure}%
\includegraphics[width=\columnwidth]{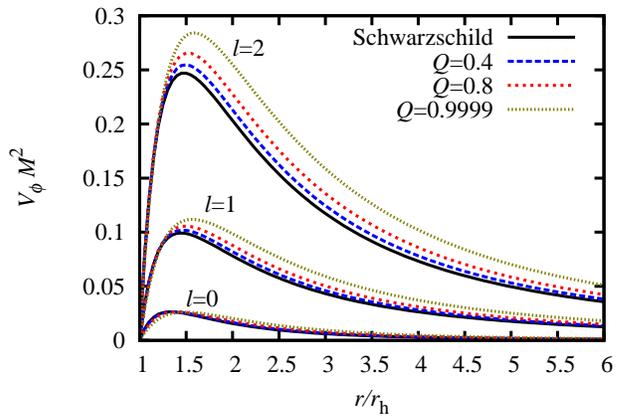}%
\caption{Scalar field potential ($V_\phi$) as function of the radial coordinate in units of the event horizon radius ($r_h$). Here we compare the Bardeen BHs with the Schwarzschild case, and we see that the shape of the potential is similar in all cases.}%
\label{fig:effectpot}%
\end{figure}
A monochromatic wave with frequency $\omega$ in a spherically symmetric background can be written as
\be
\Phi = \sum_{lm} \frac{\phi(r)}{r} Y^m_l(\theta,\varphi) e^{-i \omega t},
\label{expansion}
\ee
where $Y^m_l(\theta,\varphi)$ are the standard scalar spherical harmonics. Substituting the expansion \eqref{expansion} in Eq. \eqref{kleingordon}, and using the properties of the spherical harmonics, we get the following radial equation for $\phi(r)$:
\be
\l(-\frac{d}{dx^2}+V_\phi(r) -\omega^2\r)\phi(r)=0,\label{eq:eqr}
\ee
in which $x$ is the tortoise coordinate, defined through $dx= f(r)^{-1} dr$, and 
\be
V_\phi(r)=f\left(\frac{l (l+1)}{r^2}+\frac{f'}{r}\right)\label{eq:pot}
\ee
is the scalar field potential. Plots of $V_\phi$ for Bardeen and Schwarzschild BHs are shown in Fig. \ref{fig:effectpot}. The scalar field potential $V_\phi$ is localized, in the sense that it goes to zero at the event horizon and at infinity \cite{Fernando:2012yw}. We are interested in a solution that represents a wave coming from the past null infinity. Such a solution can be written using the so-called $in$ modes, i.e.
\be
\phi(r)\sim\l\{
\begin{array}{ll}
R_I+\mathcal{R}^{in}_{\omega l}R_I^{*}&  x\to +\infty ~(r\to +\infty),\\
  \mathcal{T}^{in}_{\omega l} R_{II} &x\to - \infty ~(r\to r_h),
\end{array}\r.
\label{eq:inmodes}
\ee
with
\bea
R_I&=&e^{-i \omega x}\sum_{j=0}^N \frac{A^j_\infty}{r^j},\\
R_{II}&=&e^{-i \omega x}\sum_{j=0}^N (r-r_h)^j A^{j}_{r_h},
\eea
where the coefficients $A^j_\infty$ and $A^j_{r_h}$ are obtained by requiring the functions $R_I$ and $R_{II}$ to be solutions of Eq. \eqref{eq:eqr} far from the BH and close to the event horizon, respectively. $|\mathcal{R}^{in}_{\omega l}|^2$ and $|\mathcal{T}^{in}_{\omega l}|^2$ are the reflection and transmission coefficients, respectively, and are related through
\be
|\mathcal{R}^{in}_{\omega l}|^2+|\mathcal{T}^{in}_{\omega l}|^2=1.
\ee

Using the solution \eqref{eq:inmodes}, the absorption cross section of planar massless scalar waves can be written as
\be
\sigma_{abs}=\sum_{l=0}^{\infty}\sigma_{l},
\label{eq:total}
\ee
with $\sigma_l$ being the partial absorption cross sections, given by
\be
\sigma_l=\frac{\pi}{\omega^2}(2l+1)\l|\mathcal{T}^{in}_{\omega l}\r|^2.
\label{eq:partialabs}
\ee

\subsection{Low- and high-frequency limits}\label{sec:lowhigh}
\begin{figure}
\includegraphics[width=\columnwidth]{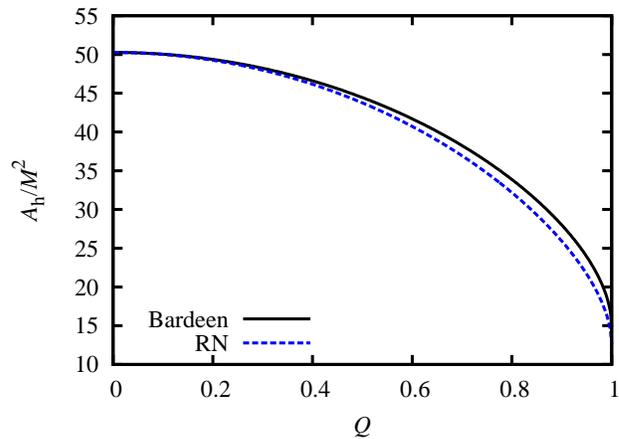}
\caption{Event horizon area of Bardeen and RN BHs as a function of the normalized charge.
}
\label{fig:area}
\end{figure}

\begin{figure}
\includegraphics[width=\columnwidth]{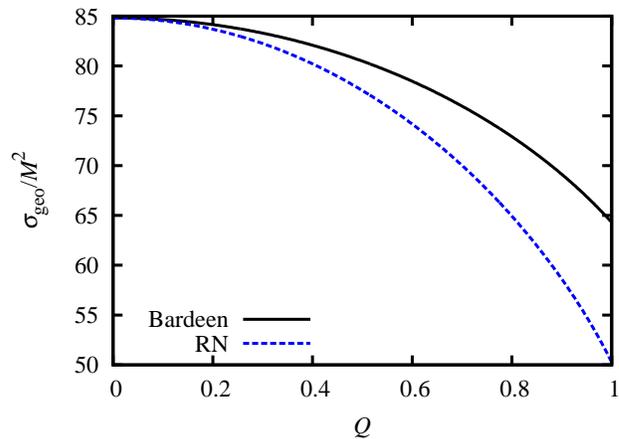}
\caption{Capture cross section of null geodesics ($\sigma_\text{geo}$) by Bardeen and RN BHs. The results for Bardeen BHs have qualitatively the same behavior of the RN BHs, with the former presenting a bigger capture cross section for the same value of the normalized charge.
}
\label{fig:capture_geo}
\end{figure}

In the low-frequency regime, it has been proven that the absorption cross section of massless scalar fields by static BHs~\cite{Das:1996we}, as well as  stationary BHs~\cite{Higuchi:2001si,Higuchi:2001sib}, tends to the area of the BH horizon. Our numerical results agree remarkably well with this low-frequency limit. In Fig. \ref{fig:area} we plot the area of the event horizon for the Bardeen and RN BHs, as a function of the normalized charge. 
We can see that the event horizon area of a Bardeen BH is bigger than the corresponding one of the RN BH with the same normalized charge. 

In the high-frequency limit, a massless scalar wave can be described by the propagation of a null vector, which follows a null geodesic. Therefore, in this limit we can consider the classical capture cross section of null geodesics to describe the absorption cross section of massless fields.

Geodesics around Bardeen BHs were also studied in Ref. \cite{Zhou:2011aa}. Here we consider null geodesics in spherically symmetric BHs. Their motion is described by the Lagrangian $L_{\text{geo}}$, that satisfies
\be
2 L_{\text{geo}} = -f(r) \dot{t}^2+f(r) \dot{r}^2+r^2 \dot{\varphi}^2=0,
\ee
in which we consider, without loss of generality, the motion in the plane $\theta =\pi/2$. The overdot indicates derivative with respect to the affine parameter of the curve. Considering the conserved quantities, namely the energy $E$ and angular momentum $L$ (see, e.g., Ref. \cite{Cardoso:2008bp}), the equation of motion becomes
\be
\dot{r}^2+L^2\frac{f(r)}{r^2}=E^2,
\ee
 which can be regarded as an energy balance equation with the effective potential 
\be
V_\text{eff}=L^2{f(r)}/{r^2}.
\label{eq:effpot}
\ee
 The high-frequency limit of the absorption cross section, also called geometric cross section, $\sigma_\text{geo}$, is then found by computing the classical capture radius of light rays in the spacetime under investigation. For spherically symmetric spacetimes, the null geodesic radius $r_l$ is obtained through $V'(r_l)=0$, with the prime denoting derivative with respect to $r$. The critical impact parameter is given by $b_c=L_c/E_c$, with $(L_c,E_c)$ being characteristic of the null circular geodesic. Therefore, we have
\be
r_l f'(r_l)-2f(r_l)=0,\label{eq:lightring}
\ee
and
\be
\sigma_{\text{geo}}=\pi b_c^2=\pi \frac{r_l^2}{f(r_l)}\label{eq:capture}.
\ee
With Eq. \eqref{eq:lightring} one finds the value of $r_l$, and by substituting this value in Eq. \eqref{eq:capture} one finds the capture (or geometric) cross section $\sigma_\text{geo}$.

In Fig. \ref{fig:capture_geo} we compare the capture cross section of the Bardeen BH with the RN BH case, as a function of the normalized charge. In general, a Bardeen BH has a bigger capture cross section, compared with the RN BH with the same value of ${Q}(>0)$.

\begin{figure}%
\includegraphics[width=\columnwidth]{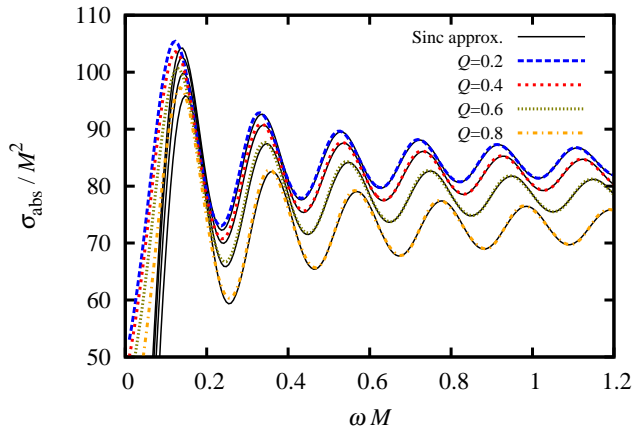}%
\caption{Comparison between the full numerical computation of the total absorption cross section of Bardeen regular BHs, given by Eq. \eqref{eq:total}, with the high-frequency (sinc) approximation, given by Eq. \eqref{eq:hfim}. We can see that the numerical and the approximate analytical results agree remarkably well even for intermediate values of the frequency.}%
\label{fig:sinc}%
\end{figure}

An improvement of the high-frequency approximation to compute the absorption cross section for spherically symmetric BHs was proposed in Ref. \cite{Decanini:2011xi}. It was shown that the oscillatory part of the absorption cross section in the eikonal limit can be written as
\be
\sigma_{\text{osc}}(\omega)=-\frac{4\Lambda_l}{\omega  \Omega_l ^2}e^{-\frac{\Lambda_l}{\Omega_l}}\sin \left(\frac{2 \pi  \omega }{\Omega_l }\right),
\label{eq:osc_abs}
\ee
where $\Lambda_l=\pi\lambda_l$, with $\lambda_l$ being the Lyapunov exponent of the null geodesic \cite{Cardoso:2008bp,Decanini:2011xi}, and $\Omega_l=d\varphi/dt=\sqrt{f(r_l)}/r_l$ being the angular velocity of the null geodesic. Therefore, we can write the high-frequency absorption cross section as
\be
\sigma_{\text{abs}}^{\text{hf}}\sim \sigma_{\text{geo}}+\sigma_{\text{osc}}.
\label{eq:hfim}
\ee
Equation~\eqref{eq:hfim} is usually referred to in the literature as the sinc approximation. In Fig. \ref{fig:sinc} we compare the results obtained through Eq. \eqref{eq:hfim} with the full numerical computation of the absorption cross section, given by Eq. \eqref{eq:total}. It is interesting to note that, although Eq. \eqref{eq:hfim} is obtained within the assumption of high frequencies, it is still a very good approximation for intermediate frequency values. 

\section{Results}\label{sec:results}

\begin{figure*}
\includegraphics[width=1.\columnwidth]{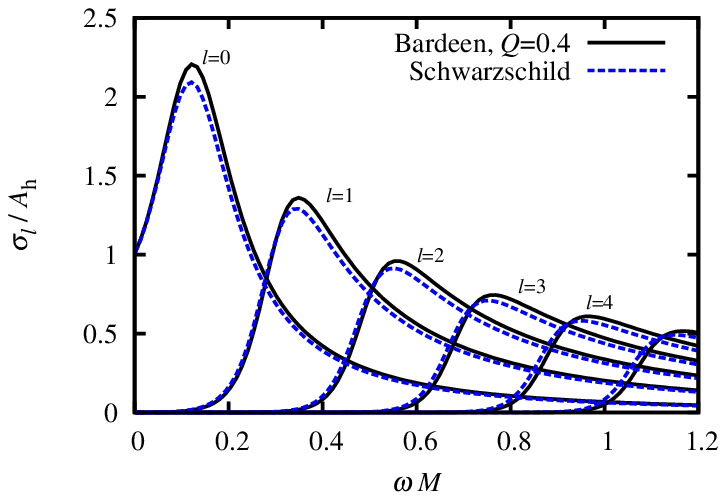}\includegraphics[width=1.\columnwidth]{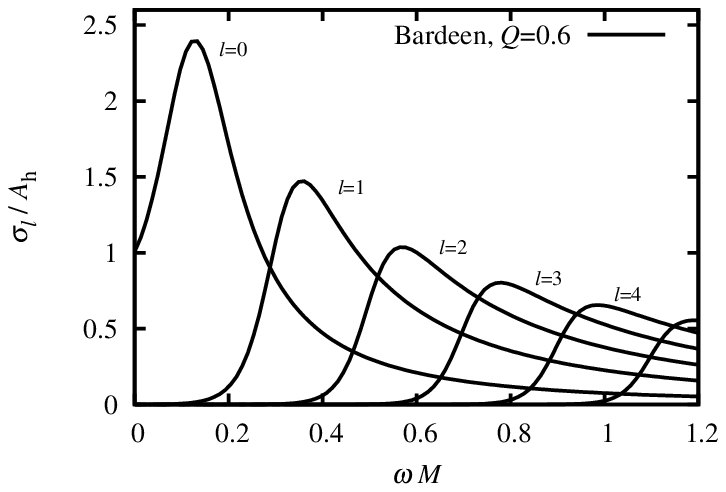}\\
\includegraphics[width=1.\columnwidth]{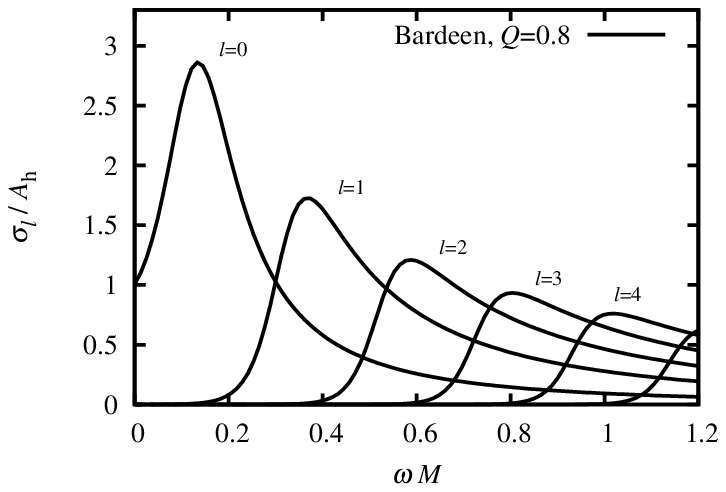}\includegraphics[width=1.\columnwidth]{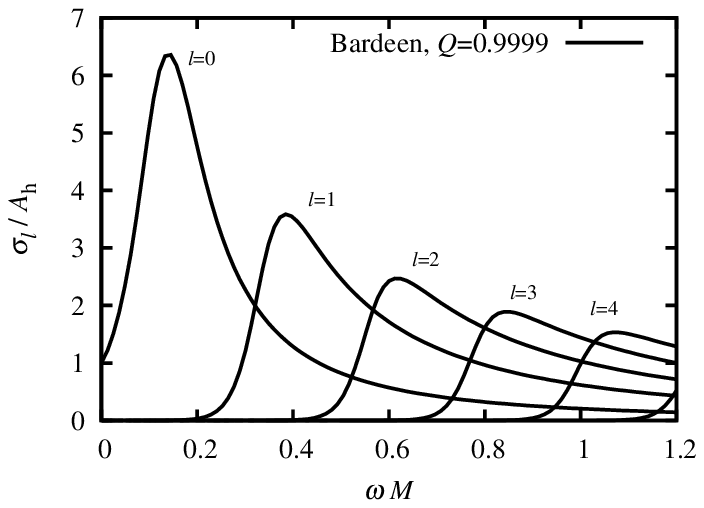}
\caption{Partial absorption cross sections of massless scalar waves by Bardeen BHs. Different frames correspond to different values of the normalized monopole charge $Q$. For comparison, in the top-left frame we also plot the partial absorption cross sections of the Schwarzschild BH (dotted lines).}
\label{fig:partialabs}
\end{figure*}

\begin{figure}
\centering
\includegraphics[width=1.\columnwidth]{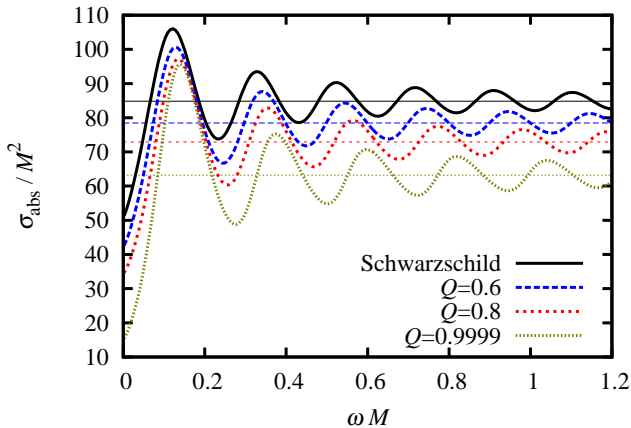} 
\caption{Absorption cross section of massless scalar waves by Bardeen BHs, compared with the capture cross section in each case (horizontal lines).
The Schwarzschild BH case is also exhibited, for comparison.}	
\label{fig:totalabs}
\end{figure}

\begin{figure}
\includegraphics[width=\columnwidth]{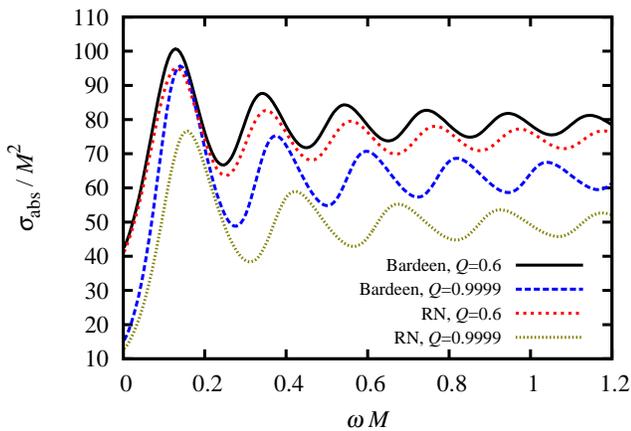} 
\caption{Comparison between the Bardeen and the RN BH cases with the same values of ${Q}$. The plots show $Q=0.6$ and $0.9999$. }
\label{fig:totalabs_rn}
\end{figure}

\begin{figure}%
\includegraphics[width=\columnwidth]{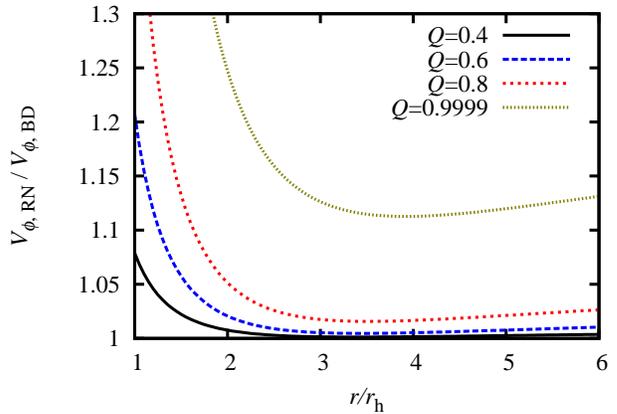}%
\caption{Ratio between the scalar field potential of the RN ($V_{\phi,\text{RN}}$) and Bardeen BHs ($V_{\phi,\text{BD}}$), for $l=0$ and $Q=0.4$, $0.6$, $0.8$, and $0.9999$. Similar results hold for higher values of $l$.}
\label{fig:potcomp}%
\end{figure}

\begin{figure}%
\includegraphics[width=\columnwidth]{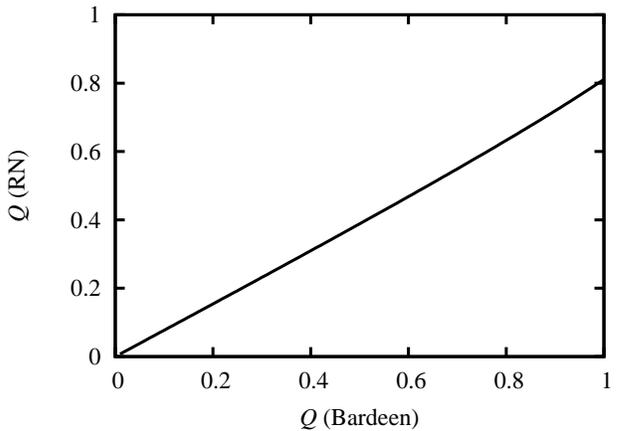}%
\caption{Values of the normalized charge for which the capture cross section of Bardeen and RN BHs are the same.}%
\label{fig:hfequiv}%
\end{figure}

\begin{figure}%
\includegraphics[width=\columnwidth]{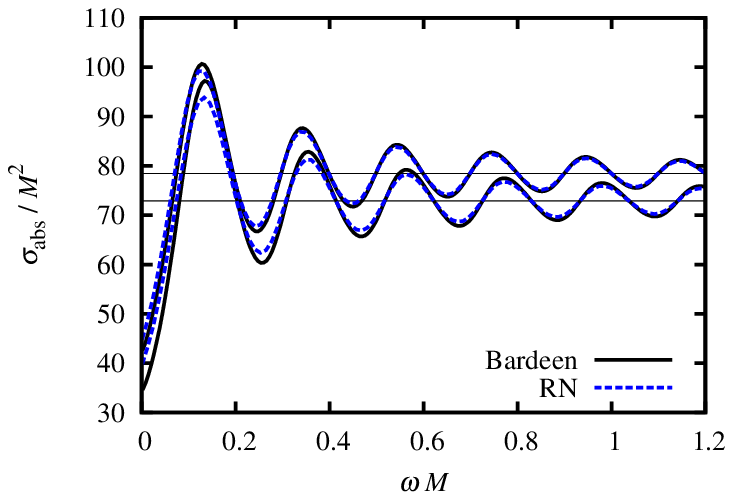}%
\caption{Total absorption cross section for Bardeen and RN BHs with the same high-frequency limit (horizontal lines). We have chosen ($Q_\text{RN},Q_\text{BD}$) to be $(0.6,0.46809)$ and $(0.8,0.63252)$. We see that their oscillation profiles are similar, but their low-frequency limits are different.}
\label{fig:hf_abs_equiv}%
\end{figure}

We have computed numerically the absorption cross section of planar massless scalar waves impinging on Bardeen BHs. In this section we show a selection of our results.

In Fig. \ref{fig:partialabs} we present the partial absorption cross sections [given by Eq. \eqref{eq:partialabs}] for ${Q}=0.4,\,0.6,\,0.8$ and $0.9999$ and for different values of $l$. We see that for $l=0$ the limit $\omega\to 0$ results in $\sigma_{\text{abs}}\to A_h$, in agreement with the result mentioned in Sec. \ref{sec:lowhigh}.

In Fig. \ref{fig:totalabs} we present the total absorption cross section [given by Eq. \eqref{eq:total}] in the Bardeen BH case, for $Q=0.6,\,0.8$ and $0.9999$, as well as in the Schwarzschild BH case. The horizontal lines are the high-frequency limits in each case. We see that the increasing of the monopole charge implies in a decreasing of the absorption cross section, in agreement with the increasing of the scattering potential (cf. Fig. \ref{fig:effectpot}), as well as with the decreasing of the horizon area (cf. Fig. \ref{fig:area}). The sum of the partial absorption cross sections generates the oscillatory profile shown in the plots of Fig. \ref{fig:totalabs}. 

In Fig. \ref{fig:totalabs_rn} we compare the absorption cross section of Bardeen and RN BHs, for the same values of $Q$. As already mentioned in Sec. \ref{sec:lowhigh} (cf. Fig. \ref{fig:capture_geo}), the high-frequency limit of the absorption cross section of the Bardeen BH is bigger than the correspondent RN BH case with the same value of $Q$. We verified that this behavior (bigger absorption for the Bardeen BH) also applies to the total absorption cross section as a whole, for any fixed value of the frequency $\omega$, for the same normalized charge $Q (>0)$. This is in accordance with the fact that the scalar field potential for the RN BH is always bigger than the corresponding one for the Bardeen BH, as it is shown in Fig. \ref{fig:potcomp}, where we plot the case in which $l=0$. Larger values of $l$ present a similar behavior.

Although for the same values of $Q$ the Bardeen BH has a bigger absorption cross section than the corresponding RN BH, for different values of the normalized charge $Q$ they can have the same capture cross section, i.e. the same high-frequency limit of the absorption cross section. In Fig. \ref{fig:hfequiv} we plot the values of the normalized charge for which the capture (or high-frequency absorption) cross section is the same for Bardeen and RN BHs. We can see from Fig. \ref{fig:hfequiv} that the RN BH must have a lower value of the normalized charge in order to have the same capture cross section of a Bardeen BH.

The equality between the high-frequency values of the absorption cross section of RN and Bardeen BHs with different normalized charges raises the following question: Can a Bardeen BH produce the same absorption spectrum of a RN BH? To answer this, we have computed the absorption cross section for configurations which have the same high-frequency limits. Some results are shown in Fig. \ref{fig:hf_abs_equiv}, where we plot the configurations for which $(Q_\text{RN},Q_\text{BD})$ are chosen to be $(0.6,0.46809)$ and $(0.8,0.63252)$. We can see that the low-frequency absorption cross section is different, although not only the high-frequency limits are the same, but also the oscillation profiles are similar. 

The similarity of the oscillation profile of Bardeen and RN BHs with the same $\sigma_\text{geo}$ can be understood as follows. From Eq. \eqref{eq:osc_abs}, we see that the oscillation pattern depends on $2\pi/\Omega_l$. Since we use configurations with the same capture cross section, the angular velocity ($\Omega_l$) of the null geodesics are also the same, once $b_c=1/\Omega_l$. Therefore, the frequency of oscillation of the two configurations will be similar.

The above scenario suggests that a regular black hole can, in principle, mimic a black hole with singularities, as far as mid-to-high-frequency absorption cross section is concerned. However, we should note that, as Fig. \ref{fig:hfequiv} shows, there is no complete correspondence between Bardeen and RN BHs absorption spectra with the same capture cross section, as it can be seen in Fig. \ref{fig:hf_abs_equiv}. Moreover, as it can be verified in Fig. \ref{fig:hfequiv}, for a Bardeen BH with $Q=1$, the corresponding RN BH with the same value of the capture cross section has a normalized charge $Q=0.8109$. Therefore, for a RN BH with a charge $Q>0.8109$ there is no correspondent Bardeen BH with the same capture cross section.

\section{Final remarks}\label{sec:final}
In this paper we presented a study of the absorption properties of regular black holes: objects which have event horizons but not singularities. For that purpose, we analyzed the case of an asymptotic planar massless scalar wave impinging upon a Bardeen regular black hole \cite{comHuang}.

We computed numerically the massless scalar absorption cross section of Bardeen regular black holes showing that the generic oscillation behavior of spherical black holes with singularities, like the Schwarzschild and Reissner--Nordstr\"om ones, is also present in the case of Bardeen regular black holes. The increasing of the monopole charge, starting from the Schwarzschild black hole case (for which $Q=0$), implies a decreasing of the absorption cross section. Our numerical results are in full agreement with the low- and high-frequency limits of the absorption cross section, which can be obtained analytically.

We compared the massless absorption cross section of a Bardeen black hole with the one of a Reissner--Nordstr\"om black hole with the same value of the normalized charge $Q$. We obtained that the behavior of the absorption cross section is qualitatively similar in both cases, but the Bardeen case always presents a bigger absorption cross section than the Reissner--Nordstr\"om case, for any fixed values of $(\omega,Q)$.	

Based on the behavior of null geodesics, we have shown that the capture cross section of a Bardeen black hole is always bigger than the corresponding one of a Reissner--Nordstr\"om black hole with the same value of $Q$. We have also shown that a Bardeen black hole can have the same capture cross section of a Reissner--Nordstr\"om black hole with a different value of $Q$. 

We computed numerically the massless scalar absorption cross section for arbitrary frequencies by Bardeen and Reissner--Nordstr\"om black holes with the same high-frequency limit. We concluded that, more than having the same capture cross section, the oscillation of the absorption cross section is similar for both cases. This comes from the fact that the oscillation depends on the angular velocity of the null circular geodesic, which is the same for the two cases. Our results suggest that some regular black holes could be mimicked by black holes with singularities, as far as mid-to-high-frequency absorption properties are concerned. The differences between the two cases manifest mainly in the low-frequency regime.

\begin{acknowledgments}
The authors would like to thank Conselho Nacional de Desenvolvimento Cient\'ifico e Tecnol\'ogico (CNPq), Coordena\c{c}\~ao de Aperfei\c{c}oamento de Pessoal de N\'ivel Superior (CAPES), and Funda\c{c}\~ao Amaz\^onia Paraense de Amparo \`a Pesquisa (FAPESPA) for partial financial support.
\end{acknowledgments}
%

\bibliography{refs_regular}

\begin{thebibliography}{10}%
\makeatletter
\providecommand \@ifxundefined [1]{%
 \ifx #1\undefined \expandafter \@firstoftwo
 \else \expandafter \@secondoftwo
\fi
}%
\providecommand \@ifnum [1]{%
 \ifnum #1\expandafter \@firstoftwo
 \else \expandafter \@secondoftwo
\fi
}%
\providecommand \enquote [1]{``#1''}%
\providecommand \bibnamefont  [1]{#1}%
\providecommand \bibfnamefont [1]{#1}%
\providecommand \citenamefont [1]{#1}%
\providecommand\href[0]{\@sanitize\@href}%
\providecommand\@href[1]{\endgroup\@@startlink{#1}\endgroup\@@href}%
\providecommand\@@href[1]{#1\@@endlink}%
\providecommand \@sanitize [0]{\begingroup\catcode`\&12\catcode`\#12\relax}%
\@ifxundefined \pdfoutput {\@firstoftwo}{%
 \@ifnum{\z@=\pdfoutput}{\@firstoftwo}{\@secondoftwo}%
}{%
 \providecommand\@@startlink[1]{\leavevmode\special{html:<a href="#1">}}%
 \providecommand\@@endlink[0]{\special{html:</a>}}%
}{%
 \providecommand\@@startlink[1]{%
  \leavevmode
  \pdfstartlink
   attr{/Border[0 0 1 ]/H/I/C[0 1 1]}%
   user{/Subtype/Link/A<</Type/Action/S/URI/URI(#1)>>}%
  \relax
 }%
 \providecommand\@@endlink[0]{\pdfendlink}%
}%
\providecommand \url  [0]{\begingroup\@sanitize \@url }%
\providecommand \@url [1]{\endgroup\@href {#1}{\urlprefix}}%
\providecommand \urlprefix [0]{URL }%
\providecommand \Eprint[0]{\href }%
\@ifxundefined \urlstyle {%
  \providecommand \doi [1]{doi:\discretionary{}{}{}#1}%
}{%
  \providecommand \doi [0]{doi:\discretionary{}{}{}\begingroup
  \urlstyle{rm}\Url }%
}%
\providecommand \doibase [0]{http://dx.doi.org/}%
\providecommand \Doi[1]{\href{\doibase#1}}%
\providecommand \bibAnnote [3]{%
  \BibitemShut{#1}%
  \begin{quotation}\noindent
    \textsc{Key:}\ #2\\\textsc{Annotation:}\ #3%
  \end{quotation}%
}%
\providecommand \bibAnnoteFile [2]{%
  \IfFileExists{#2}{\bibAnnote {#1} {#2} {\input{#2}}}{}%
}%
\providecommand \typeout [0]{\immediate \write \m@ne }%
\providecommand \selectlanguage [0]{\@gobble}%
\providecommand \bibinfo [0]{\@secondoftwo}%
\providecommand \bibfield [0]{\@secondoftwo}%
\providecommand \translation [1]{[#1]}%
\providecommand \BibitemOpen[0]{}%
\providecommand \bibitemStop [0]{}%
\providecommand \bibitemNoStop [0]{.\EOS\space}%
\providecommand \EOS [0]{\spacefactor3000\relax}%
\providecommand \BibitemShut [1]{\csname bibitem#1\endcsname}%
\bibitem{Narayan:2005ie}%
  \BibitemOpen
  \bibfield{author}{%
  \bibinfo {author} {\bibfnamefont{R.}~\bibnamefont{Narayan}},\ }%
  \bibfield{journal}{%
  \Doi{10.1088/1367-2630/7/1/199}{\bibinfo {journal} {New J. Phys.}}\ }%
  \textbf{\bibinfo {volume} {7}},\ \bibinfo {pages} {199} (\bibinfo {year}
  {2005}),\ \Eprint{http://arxiv.org/abs/gr-qc/0506078}{arXiv:gr-qc/0506078
  [gr-qc]}%
  \bibAnnoteFile{NoStop}{Narayan:2005ie}%
\bibitem{heuslerreview}%
  \BibitemOpen
  \bibfield{author}{%
  \bibinfo {author} {\bibfnamefont{M.}~\bibnamefont{Heusler}},\ }%
  \bibfield{journal}{%
  \bibinfo {journal} {Living Rev. Relativity}\ }%
  \textbf{\bibinfo {volume} {1}},\ \bibinfo {pages} {6} (\bibinfo {year}
  {1998}),\ \url{http://www.livingreviews.org/lrr-1998-6}%
  \bibAnnoteFile{NoStop}{heuslerreview}%
\bibitem{Penrose:1969pc}%
  \BibitemOpen
  \bibfield{author}{%
  \bibinfo {author} {\bibfnamefont{R.}~\bibnamefont{Penrose}},\ }%
  \bibfield{journal}{%
  \bibinfo {journal} {Riv.Nuovo Cim.}\ }%
  \textbf{\bibinfo {volume} {1}},\ \bibinfo {pages} {252} (\bibinfo {year}
  {1969})%
  \bibAnnoteFile{NoStop}{Penrose:1969pc}%
\bibitem{Wald:1997wa}%
  \BibitemOpen
  \bibfield{author}{%
  \bibinfo {author} {\bibfnamefont{R.~M.}\ \bibnamefont{Wald}}\ }%
  \Eprint{http://arxiv.org/abs/gr-qc/9710068}{arXiv:gr-qc/9710068 [gr-qc]}%
  \bibAnnoteFile{NoStop}{Wald:1997wa}%
\bibitem{hawkingb}%
  \BibitemOpen
  \bibfield{author}{%
  \bibinfo {author} {\bibfnamefont{S.~W.}\ \bibnamefont{Hawking}}\ and\
  \bibinfo {author} {\bibfnamefont{G.~F.~R.}\ \bibnamefont{Ellis}},\ }%
  \emph{\bibinfo {title} {{The Large Scale Structure of Space-Time}}}\
  (\bibinfo {publisher} {Cambridge University Press},\ \bibinfo {address}
  {Cambridge, England},\ \bibinfo {year} {1973})%
  \bibAnnoteFile{NoStop}{hawkingb}%
\bibitem{Clifton:2011jh}%
  \BibitemOpen
  \bibfield{author}{%
  \bibinfo {author} {\bibfnamefont{T.}~\bibnamefont{Clifton}}, \bibinfo
  {author} {\bibfnamefont{P.~G.}\ \bibnamefont{Ferreira}}, \bibinfo {author}
  {\bibfnamefont{A.}~\bibnamefont{Padilla}},\ and\ \bibinfo {author}
  {\bibfnamefont{C.}~\bibnamefont{Skordis}},\ }%
  \bibfield{journal}{%
  \Doi{10.1016/j.physrep.2012.01.001}{\bibinfo {journal} {Phys. Rep.}}\ }%
  \textbf{\bibinfo {volume} {513}},\ \bibinfo {pages} {1} (\bibinfo {year}
  {2012}),\ \Eprint{http://arxiv.org/abs/1106.2476}{arXiv:1106.2476
  [astro-ph.CO]}%
  \bibAnnoteFile{NoStop}{Clifton:2011jh}%
\bibitem{Bardeen}%
  \BibitemOpen
  \bibfield{author}{%
  \bibinfo {author} {\bibfnamefont{J.}~\bibnamefont{Bardeen}},\ }%
  \emph{\bibinfo {title} {{Proceedings of GR 5, Tbilisi, USSR, 1968}}}\
  (\bibinfo {publisher} {unpublished})%
  \bibAnnoteFile{NoStop}{Bardeen}%
\bibitem{AyonBeato:2000zs}%
  \BibitemOpen
  \bibfield{author}{%
  \bibinfo {author} {\bibfnamefont{E.}~\bibnamefont{Ayon-Beato}}\ and\ \bibinfo
  {author} {\bibfnamefont{A.}~\bibnamefont{Garcia}},\ }%
  \bibfield{journal}{%
  \Doi{10.1016/S0370-2693(00)01125-4}{\bibinfo {journal} {Phys. Lett. B}}\ }%
  \textbf{\bibinfo {volume} {493}},\ \bibinfo {pages} {149} (\bibinfo {year}
  {2000}),\ \Eprint{http://arxiv.org/abs/gr-qc/0009077}{arXiv:gr-qc/0009077
  [gr-qc]}%
  \bibAnnoteFile{NoStop}{AyonBeato:2000zs}%
\bibitem{Ansoldi:2008jw}%
  \BibitemOpen
  \bibfield{author}{%
  \bibinfo {author} {\bibfnamefont{S.}~\bibnamefont{Ansoldi}}\ }%
  \Eprint{http://arxiv.org/abs/0802.0330}{arXiv:0802.0330 [gr-qc]}%
  \bibAnnoteFile{NoStop}{Ansoldi:2008jw}%
\bibitem{Lemos:2011dq}%
  \BibitemOpen
  \bibfield{author}{%
  \bibinfo {author} {\bibfnamefont{J.~P.~S.}\ \bibnamefont{Lemos}}\ and\
  \bibinfo {author} {\bibfnamefont{V.~T.}\ \bibnamefont{Zanchin}},\ }%
  \bibfield{journal}{%
  \Doi{10.1103/PhysRevD.83.124005}{\bibinfo {journal} {Phys. Rev. D}}\ }%
  \textbf{\bibinfo {volume} {83}},\ \bibinfo {pages} {124005} (\bibinfo {year}
  {2011}),\ \Eprint{http://arxiv.org/abs/1104.4790}{arXiv:1104.4790 [gr-qc]}%
  \bibAnnoteFile{NoStop}{Lemos:2011dq}%
\bibitem{Nollert:1999ji}%
  \BibitemOpen
  \bibfield{author}{%
  \bibinfo {author} {\bibfnamefont{H.-P.}\ \bibnamefont{Nollert}},\ }%
  \bibfield{journal}{%
  \Doi{10.1088/0264-9381/16/12/201}{\bibinfo {journal} {Classical Quantum
  Gravity}}\ }%
  \textbf{\bibinfo {volume} {16}},\ \bibinfo {pages} {R159} (\bibinfo {year}
  {1999})%
  \bibAnnoteFile{NoStop}{Nollert:1999ji}%
\bibitem{Berti:2009kk}%
  \BibitemOpen
  \bibfield{author}{%
  \bibinfo {author} {\bibfnamefont{E.}~\bibnamefont{Berti}}, \bibinfo {author}
  {\bibfnamefont{V.}~\bibnamefont{Cardoso}},\ and\ \bibinfo {author}
  {\bibfnamefont{A.~O.}\ \bibnamefont{Starinets}},\ }%
  \bibfield{journal}{%
  \Doi{10.1088/0264-9381/26/16/163001}{\bibinfo {journal} {Classical Quantum
  Gravity}}\ }%
  \textbf{\bibinfo {volume} {26}},\ \bibinfo {pages} {163001} (\bibinfo {year}
  {2009}),\ \Eprint{http://arxiv.org/abs/0905.2975}{arXiv:0905.2975 [gr-qc]}%
  \bibAnnoteFile{NoStop}{Berti:2009kk}%
\bibitem{Flachi:2012nv}%
  \BibitemOpen
  \bibfield{author}{%
  \bibinfo {author} {\bibfnamefont{A.}~\bibnamefont{Flachi}}\ and\ \bibinfo
  {author} {\bibfnamefont{J.~P.~S.}\ \bibnamefont{Lemos}},\ }%
  \bibfield{journal}{%
  \Doi{10.1103/PhysRevD.87.024034}{\bibinfo {journal} {Phys. Rev. D}}\ }%
  \textbf{\bibinfo {volume} {87}},\ \bibinfo {pages} {024034} (\bibinfo {year}
  {2013}),\ \Eprint{http://arxiv.org/abs/1211.6212}{arXiv:1211.6212 [gr-qc]}%
  \bibAnnoteFile{NoStop}{Flachi:2012nv}%
\bibitem{Li:2013fka}%
  \BibitemOpen
  \bibfield{author}{%
  \bibinfo {author} {\bibfnamefont{J.}~\bibnamefont{Li}}, \bibinfo {author}
  {\bibfnamefont{M.}~\bibnamefont{Hong}},\ and\ \bibinfo {author}
  {\bibfnamefont{K.}~\bibnamefont{Lin}},\ }%
  \bibfield{journal}{%
  \Doi{10.1103/PhysRevD.88.064001}{\bibinfo {journal} {Phys. Rev. D}}\ }%
  \textbf{\bibinfo {volume} {88}},\ \bibinfo {pages} {064001} (\bibinfo {year}
  {2013}),\ \Eprint{http://arxiv.org/abs/1308.6499}{arXiv:1308.6499 [gr-qc]}%
  \bibAnnoteFile{NoStop}{Li:2013fka}%
\bibitem{Lin:2013ofa}%
  \BibitemOpen
  \bibfield{author}{%
  \bibinfo {author} {\bibfnamefont{K.}~\bibnamefont{Lin}}, \bibinfo {author}
  {\bibfnamefont{J.}~\bibnamefont{Li}},\ and\ \bibinfo {author}
  {\bibfnamefont{S.}~\bibnamefont{Yang}},\ }%
  \bibfield{journal}{%
  \Doi{10.1007/s10773-013-1682-4}{\bibinfo {journal} {Int. J. Theor. Phys.}}\
  }%
  \textbf{\bibinfo {volume} {52}},\ \bibinfo {pages} {3771} (\bibinfo {year}
  {2013})%
  \bibAnnoteFile{NoStop}{Lin:2013ofa}%
\bibitem{Vishveshwara:1970zz}%
  \BibitemOpen
  \bibfield{author}{%
  \bibinfo {author} {\bibfnamefont{C.}~\bibnamefont{Vishveshwara}},\ }%
  \bibfield{journal}{%
  \Doi{10.1038/227936a0}{\bibinfo {journal} {Nature}}\ }%
  \textbf{\bibinfo {volume} {227}},\ \bibinfo {pages} {936} (\bibinfo {year}
  {1970})%
  \bibAnnoteFile{NoStop}{Vishveshwara:1970zz}%
\bibitem{Andersson:1996cm}%
  \BibitemOpen
  \bibfield{author}{%
  \bibinfo {author} {\bibfnamefont{N.}~\bibnamefont{Andersson}},\ }%
  \bibfield{journal}{%
  \Doi{10.1103/PhysRevD.55.468}{\bibinfo {journal} {Phys. Rev. D}}\ }%
  \textbf{\bibinfo {volume} {55}},\ \bibinfo {pages} {468} (\bibinfo {year}
  {1997}),\ \Eprint{http://arxiv.org/abs/gr-qc/9607064}{arXiv:gr-qc/9607064
  [gr-qc]}%
  \bibAnnoteFile{NoStop}{Andersson:1996cm}%
\bibitem{Granato:2003ch}%
  \BibitemOpen
  \bibfield{author}{%
  \bibinfo {author} {\bibfnamefont{G.~L.}\ \bibnamefont{Granato}}, \bibinfo
  {author} {\bibfnamefont{G.}~\bibnamefont{De~Zotti}}, \bibinfo {author}
  {\bibfnamefont{L.}~\bibnamefont{Silva}}, \bibinfo {author}
  {\bibfnamefont{A.}~\bibnamefont{Bressan}},\ and\ \bibinfo {author}
  {\bibfnamefont{L.}~\bibnamefont{Danese}},\ }%
  \bibfield{journal}{%
  \Doi{10.1086/379875}{\bibinfo {journal} {Astrophys. J.}}\ }%
  \textbf{\bibinfo {volume} {600}},\ \bibinfo {pages} {580} (\bibinfo {year}
  {2004}),\
  \Eprint{http://arxiv.org/abs/astro-ph/0307202}{arXiv:astro-ph/0307202
  [astro-ph]}%
  \bibAnnoteFile{NoStop}{Granato:2003ch}%
\bibitem{Marconi:2003tg}%
  \BibitemOpen
  \bibfield{author}{%
  \bibinfo {author} {\bibfnamefont{A.}~\bibnamefont{Marconi}}, \bibinfo
  {author} {\bibfnamefont{G.}~\bibnamefont{Risaliti}}, \bibinfo {author}
  {\bibfnamefont{R.}~\bibnamefont{Gilli}}, \bibinfo {author}
  {\bibfnamefont{L.}~\bibnamefont{Hunt}}, \bibinfo {author}
  {\bibfnamefont{R.}~\bibnamefont{Maiolino}}, \emph{et~al.},\ }%
  \bibfield{journal}{%
  \Doi{10.1111/j.1365-2966.2004.07765.x}{\bibinfo {journal} {Mon. Not. R.
  Astron. Soc.}}\ }%
  \textbf{\bibinfo {volume} {351}},\ \bibinfo {pages} {169} (\bibinfo {year}
  {2004}),\
  \Eprint{http://arxiv.org/abs/astro-ph/0311619}{arXiv:astro-ph/0311619
  [astro-ph]}%
  \bibAnnoteFile{NoStop}{Marconi:2003tg}%
\bibitem{Ferrarese:2004qr}%
  \BibitemOpen
  \bibfield{author}{%
  \bibinfo {author} {\bibfnamefont{L.}~\bibnamefont{Ferrarese}}\ and\ \bibinfo
  {author} {\bibfnamefont{H.}~\bibnamefont{Ford}},\ }%
  \bibfield{journal}{%
  \Doi{10.1007/s11214-005-3947-6}{\bibinfo {journal} {Space Sci.Rev.}}\ }%
  \textbf{\bibinfo {volume} {116}},\ \bibinfo {pages} {523} (\bibinfo {year}
  {2005}),\
  \Eprint{http://arxiv.org/abs/astro-ph/0411247}{arXiv:astro-ph/0411247
  [astro-ph]}%
  \bibAnnoteFile{NoStop}{Ferrarese:2004qr}%
\bibitem{Matzner:1968}%
  \BibitemOpen
  \bibfield{author}{%
  \bibinfo {author} {\bibfnamefont{R.~A.}\ \bibnamefont{Matzner}},\ }%
  \bibfield{journal}{%
  \Doi{10.1063/1.1664470}{\bibinfo {journal} {J. of Math. Phys.}}\ }%
  \textbf{\bibinfo {volume} {9}},\ \bibinfo {pages} {163} (\bibinfo {year}
  {1968})%
  \bibAnnoteFile{NoStop}{Matzner:1968}%
\bibitem{Sanchez:1977si}%
  \BibitemOpen
  \bibfield{author}{%
  \bibinfo {author} {\bibfnamefont{N.~G.}\ \bibnamefont{Sanchez}},\ }%
  \bibfield{journal}{%
  \Doi{10.1103/PhysRevD.18.1030}{\bibinfo {journal} {Phys. Rev. D}}\ }%
  \textbf{\bibinfo {volume} {18}},\ \bibinfo {pages} {1030} (\bibinfo {year}
  {1978})%
  \bibAnnoteFile{NoStop}{Sanchez:1977si}%
\bibitem{Futterman:1988ni}%
  \BibitemOpen
  \bibfield{author}{%
  \bibinfo {author} {\bibfnamefont{J.~A.~H.}\ \bibnamefont{{Futterman}}},
  \bibinfo {author} {\bibfnamefont{F.~A.}\ \bibnamefont{{Handler}}},\ and\
  \bibinfo {author} {\bibfnamefont{R.~A.}\ \bibnamefont{{Matzner}}},\ }%
  \emph{\bibinfo {title} {{Scattering from black holes}}}\ (\bibinfo
  {publisher} {Cambridge University Press},\ \bibinfo {address} {Cambridge,
  England},\ \bibinfo {year} {1988})%
  \bibAnnoteFile{NoStop}{Futterman:1988ni}%
\bibitem{Higuchi:2001si}%
  \BibitemOpen
  \bibfield{author}{%
  \bibinfo {author} {\bibfnamefont{A.}~\bibnamefont{Higuchi}},\ }%
  \bibfield{journal}{%
  \Doi{10.1088/0264-9381/18/20/102}{\bibinfo {journal} {Classical Quantum
  Gravity}}\ }%
  \textbf{\bibinfo {volume} {18}},\ \bibinfo {pages} {L139} (\bibinfo {year}
  {2001}),\ \Eprint{http://arxiv.org/abs/hep-th/0108144}{arXiv:hep-th/0108144
  [hep-th]}%
  \bibAnnoteFile{NoStop}{Higuchi:2001si}%
\bibitem{Higuchi:2001sib}%
  \BibitemOpen
  \bibfield{author}{%
  \bibinfo {author} {\bibfnamefont{A.}~\bibnamefont{Higuchi}},\ }%
  \bibfield{journal}{%
  \bibinfo {journal} {Classical Quantum Gravity}\ }%
  \textbf{\bibinfo {volume} {19}},\ \bibinfo {pages} {599} (\bibinfo {year}
  {2002}),\ \url{http://stacks.iop.org/0264-9381/19/i=3/a=401}%
  \bibAnnoteFile{NoStop}{Higuchi:2001sib}%
\bibitem{Jung:2005mr}%
  \BibitemOpen
  \bibfield{author}{%
  \bibinfo {author} {\bibfnamefont{E.}~\bibnamefont{Jung}}\ and\ \bibinfo
  {author} {\bibfnamefont{D.}~\bibnamefont{Park}},\ }%
  \bibfield{journal}{%
  \Doi{10.1016/j.nuclphysb.2005.03.037}{\bibinfo {journal} {Nucl.Phys.}}\ }%
  \textbf{\bibinfo {volume} {B717}},\ \bibinfo {pages} {272} (\bibinfo {year}
  {2005}),\ \Eprint{http://arxiv.org/abs/hep-th/0502002}{arXiv:hep-th/0502002
  [hep-th]}%
  \bibAnnoteFile{NoStop}{Jung:2005mr}%
\bibitem{Macedo:2013afa}%
  \BibitemOpen
  \bibfield{author}{%
  \bibinfo {author} {\bibfnamefont{C.~F.~B.}\ \bibnamefont{Macedo}}, \bibinfo
  {author} {\bibfnamefont{L.~C.~S.}\ \bibnamefont{Leite}}, \bibinfo {author}
  {\bibfnamefont{E.~S.}\ \bibnamefont{Oliveira}}, \bibinfo {author}
  {\bibfnamefont{S.~R.}\ \bibnamefont{Dolan}},\ and\ \bibinfo {author}
  {\bibfnamefont{L.~C.~B.}\ \bibnamefont{Crispino}},\ }%
  \bibfield{journal}{%
  \Doi{10.1103/PhysRevD.88.064033}{\bibinfo {journal} {Phys. Rev. D}}\ }%
  \textbf{\bibinfo {volume} {88}},\ \bibinfo {pages} {064033} (\bibinfo {year}
  {2013}),\ \Eprint{http://arxiv.org/abs/1308.0018}{arXiv:1308.0018 [gr-qc]}%
  \bibAnnoteFile{NoStop}{Macedo:2013afa}%
\bibitem{Benone:2014qaa}%
  \BibitemOpen
  \bibfield{author}{%
  \bibinfo {author} {\bibfnamefont{C.~L.}\ \bibnamefont{Benone}}, \bibinfo
  {author} {\bibfnamefont{E.~S.}\ \bibnamefont{de~Oliveira}}, \bibinfo {author}
  {\bibfnamefont{S.~R.}\ \bibnamefont{Dolan}},\ and\ \bibinfo {author}
  {\bibfnamefont{L.~C.~B.}\ \bibnamefont{Crispino}},\ }%
  \bibfield{journal}{%
  \Doi{10.1103/PhysRevD.89.104053}{\bibinfo {journal} {Phys. Rev. D}}\ }%
  \textbf{\bibinfo {volume} {89}},\ \bibinfo {pages} {104053} (\bibinfo {year}
  {2014}),\ \Eprint{http://arxiv.org/abs/1404.0687}{arXiv:1404.0687 [gr-qc]}%
  \bibAnnoteFile{NoStop}{Benone:2014qaa}%
\bibitem{Huang:2007us}%
  \BibitemOpen
  \bibfield{author}{%
  \bibinfo {author} {\bibfnamefont{L.}~\bibnamefont{Huang}}, \bibinfo {author}
  {\bibfnamefont{M.}~\bibnamefont{Cai}}, \bibinfo {author}
  {\bibfnamefont{Z.-Q.}\ \bibnamefont{Shen}},\ and\ \bibinfo {author}
  {\bibfnamefont{F.}~\bibnamefont{Yuan}},\ }%
  \bibfield{journal}{%
  \Doi{10.1111/j.1365-2966.2007.11713.x}{\bibinfo {journal} {Mon. Not. R.
  Astron. Soc.}}\ }%
  \textbf{\bibinfo {volume} {379}},\ \bibinfo {pages} {833} (\bibinfo {year}
  {2007}),\
  \Eprint{http://arxiv.org/abs/astro-ph/0703254}{arXiv:astro-ph/0703254
  [ASTRO-PH]}%
  \bibAnnoteFile{NoStop}{Huang:2007us}%
\bibitem{Johannsen:2010ru}%
  \BibitemOpen
  \bibfield{author}{%
  \bibinfo {author} {\bibfnamefont{T.}~\bibnamefont{Johannsen}}\ and\ \bibinfo
  {author} {\bibfnamefont{D.}~\bibnamefont{Psaltis}},\ }%
  \bibfield{journal}{%
  \Doi{10.1088/0004-637X/718/1/446}{\bibinfo {journal} {Astrophys. J.}}\ }%
  \textbf{\bibinfo {volume} {718}},\ \bibinfo {pages} {446} (\bibinfo {year}
  {2010}),\ \Eprint{http://arxiv.org/abs/1005.1931}{arXiv:1005.1931
  [astro-ph.HE]}%
  \bibAnnoteFile{NoStop}{Johannsen:2010ru}%
\bibitem{Li:2013jra}%
  \BibitemOpen
  \bibfield{author}{%
  \bibinfo {author} {\bibfnamefont{Z.}~\bibnamefont{Li}}\ and\ \bibinfo
  {author} {\bibfnamefont{C.}~\bibnamefont{Bambi}},\ }%
  \bibfield{journal}{%
  \Doi{10.1088/1475-7516/2014/01/041}{\bibinfo {journal} {J. Cosmol. Astropart.
  Phys. 01}}\ }%
  \textbf{\bibinfo {volume} {1401}},\ \bibinfo {pages} {041} (\bibinfo {year}
  {2014}),\ \Eprint{http://arxiv.org/abs/1309.1606}{arXiv:1309.1606 [gr-qc]}%
  \bibAnnoteFile{NoStop}{Li:2013jra}%
\bibitem{Petrich:1988zz}%
  \BibitemOpen
  \bibfield{author}{%
  \bibinfo {author} {\bibfnamefont{L.~I.}\ \bibnamefont{Petrich}}, \bibinfo
  {author} {\bibfnamefont{S.~L.}\ \bibnamefont{Shapiro}},\ and\ \bibinfo
  {author} {\bibfnamefont{S.~A.}\ \bibnamefont{Teukolsky}},\ }%
  \bibfield{journal}{%
  \Doi{10.1103/PhysRevLett.60.1781}{\bibinfo {journal} {Phys.Rev.Lett.}}\ }%
  \textbf{\bibinfo {volume} {60}},\ \bibinfo {pages} {1781} (\bibinfo {year}
  {1988})%
  \bibAnnoteFile{NoStop}{Petrich:1988zz}%
\bibitem{Macedo:2013qea}%
  \BibitemOpen
  \bibfield{author}{%
  \bibinfo {author} {\bibfnamefont{C.~F.~B.}\ \bibnamefont{Macedo}}, \bibinfo
  {author} {\bibfnamefont{P.}~\bibnamefont{Pani}}, \bibinfo {author}
  {\bibfnamefont{V.}~\bibnamefont{Cardoso}},\ and\ \bibinfo {author}
  {\bibfnamefont{L.~C.~B.}\ \bibnamefont{Crispino}},\ }%
  \bibfield{journal}{%
  \Doi{10.1088/0004-637X/774/1/48}{\bibinfo {journal} {Astrophys. J.}}\ }%
  \textbf{\bibinfo {volume} {774}},\ \bibinfo {pages} {48} (\bibinfo {year}
  {2013}),\ \Eprint{http://arxiv.org/abs/1302.2646}{arXiv:1302.2646}%
  \bibAnnoteFile{NoStop}{Macedo:2013qea}%
\bibitem{Macedo:2013jja}%
  \BibitemOpen
  \bibfield{author}{%
  \bibinfo {author} {\bibfnamefont{C.~F.~B.}\ \bibnamefont{Macedo}}, \bibinfo
  {author} {\bibfnamefont{P.}~\bibnamefont{Pani}}, \bibinfo {author}
  {\bibfnamefont{V.}~\bibnamefont{Cardoso}},\ and\ \bibinfo {author}
  {\bibfnamefont{L.~C.~B.}\ \bibnamefont{Crispino}},\ }%
  \bibfield{journal}{%
  \Doi{10.1103/PhysRevD.88.064046}{\bibinfo {journal} {Phys. Rev. D}}\ }%
  \textbf{\bibinfo {volume} {88}},\ \bibinfo {pages} {064046} (\bibinfo {year}
  {2013}),\ \Eprint{http://arxiv.org/abs/1307.4812}{arXiv:1307.4812 [gr-qc]}%
  \bibAnnoteFile{NoStop}{Macedo:2013jja}%
\bibitem{PhysRevD.79.064022}%
  \BibitemOpen
  \bibfield{author}{%
  \bibinfo {author} {\bibfnamefont{L.~C.~B.}\ \bibnamefont{Crispino}}, \bibinfo
  {author} {\bibfnamefont{S.~R.}\ \bibnamefont{Dolan}},\ and\ \bibinfo {author}
  {\bibfnamefont{E.~S.}\ \bibnamefont{Oliveira}},\ }%
  \bibfield{journal}{%
  \Doi{10.1103/PhysRevD.79.064022}{\bibinfo {journal} {Phys. Rev. D}}\ }%
  \textbf{\bibinfo {volume} {79}},\ \bibinfo {pages} {064022} (\bibinfo {month}
  {Mar}\ \bibinfo {year} {2009}),\
  \url{http://link.aps.org/doi/10.1103/PhysRevD.79.064022}%
  \bibAnnoteFile{NoStop}{PhysRevD.79.064022}%
\bibitem{Fernando:2012yw}%
  \BibitemOpen
  \bibfield{author}{%
  \bibinfo {author} {\bibfnamefont{S.}~\bibnamefont{Fernando}}\ and\ \bibinfo
  {author} {\bibfnamefont{J.}~\bibnamefont{Correa}},\ }%
  \bibfield{journal}{%
  \Doi{10.1103/PhysRevD.86.064039}{\bibinfo {journal} {Phys. Rev. D}}\ }%
  \textbf{\bibinfo {volume} {86}},\ \bibinfo {pages} {064039} (\bibinfo {year}
  {2012}),\ \Eprint{http://arxiv.org/abs/1208.5442}{arXiv:1208.5442 [gr-qc]}%
  \bibAnnoteFile{NoStop}{Fernando:2012yw}%
\bibitem{Das:1996we}%
  \BibitemOpen
  \bibfield{author}{%
  \bibinfo {author} {\bibfnamefont{S.~R.}\ \bibnamefont{Das}}, \bibinfo
  {author} {\bibfnamefont{G.~W.}\ \bibnamefont{Gibbons}},\ and\ \bibinfo
  {author} {\bibfnamefont{S.~D.}\ \bibnamefont{Mathur}},\ }%
  \bibfield{journal}{%
  \Doi{10.1103/PhysRevLett.78.417}{\bibinfo {journal} {Phys.Rev.Lett.}}\ }%
  \textbf{\bibinfo {volume} {78}},\ \bibinfo {pages} {417} (\bibinfo {year}
  {1997}),\ \Eprint{http://arxiv.org/abs/hep-th/9609052}{arXiv:hep-th/9609052
  [hep-th]}%
  \bibAnnoteFile{NoStop}{Das:1996we}%
\bibitem{Zhou:2011aa}%
  \BibitemOpen
  \bibfield{author}{%
  \bibinfo {author} {\bibfnamefont{S.}~\bibnamefont{Zhou}}, \bibinfo {author}
  {\bibfnamefont{J.}~\bibnamefont{Chen}},\ and\ \bibinfo {author}
  {\bibfnamefont{Y.}~\bibnamefont{Wang}},\ }%
  \bibfield{journal}{%
  \Doi{10.1142/S0218271812500770}{\bibinfo {journal} {Int. J. Mod. Phys. D}}\
  }%
  \textbf{\bibinfo {volume} {21}},\ \bibinfo {pages} {1250077} (\bibinfo {year}
  {2012}),\ \Eprint{http://arxiv.org/abs/1112.5909}{arXiv:1112.5909 [gr-qc]}%
  \bibAnnoteFile{NoStop}{Zhou:2011aa}%
\bibitem{Cardoso:2008bp}%
  \BibitemOpen
  \bibfield{author}{%
  \bibinfo {author} {\bibfnamefont{V.}~\bibnamefont{Cardoso}}, \bibinfo
  {author} {\bibfnamefont{A.~S.}\ \bibnamefont{Miranda}}, \bibinfo {author}
  {\bibfnamefont{E.}~\bibnamefont{Berti}}, \bibinfo {author}
  {\bibfnamefont{H.}~\bibnamefont{Witek}},\ and\ \bibinfo {author}
  {\bibfnamefont{V.~T.}\ \bibnamefont{Zanchin}},\ }%
  \bibfield{journal}{%
  \Doi{10.1103/PhysRevD.79.064016}{\bibinfo {journal} {Phys. Rev. D}}\ }%
  \textbf{\bibinfo {volume} {79}},\ \bibinfo {pages} {064016} (\bibinfo {year}
  {2009}),\ \Eprint{http://arxiv.org/abs/0812.1806}{arXiv:0812.1806 [hep-th]}%
  \bibAnnoteFile{NoStop}{Cardoso:2008bp}%
\bibitem{Decanini:2011xi}%
  \BibitemOpen
  \bibfield{author}{%
  \bibinfo {author} {\bibfnamefont{Y.}~\bibnamefont{Decanini}}, \bibinfo
  {author} {\bibfnamefont{G.}~\bibnamefont{Esposito-Farese}},\ and\ \bibinfo
  {author} {\bibfnamefont{A.}~\bibnamefont{Folacci}},\ }%
  \bibfield{journal}{%
  \Doi{10.1103/PhysRevD.83.044032}{\bibinfo {journal} {Phys. Rev. D}}\ }%
  \textbf{\bibinfo {volume} {83}},\ \bibinfo {pages} {044032} (\bibinfo {year}
  {2011}),\ \Eprint{http://arxiv.org/abs/1101.0781}{arXiv:1101.0781 [gr-qc]}%
  \bibAnnoteFile{NoStop}{Decanini:2011xi}%
\bibitem{comHuang}%
  \BibitemOpen
  \bibinfo {note} {After this work was basically concluded, we became aware of
  an attempt to compute the scalar absorption cross section of regular black
  holes in Ref.~\cite{Huang2014}. The authors of Ref.~\cite{Huang2014} used the
  Wentzel–Kramers–Brillouin (WKB) method with the P\"oschl--Teller
  potential to solve the differential Eq. \eqref{eq:eqr}. However, their
  results do not seem correct, and they differ considerably from ours. In
  particular, the authors of Ref.~\cite{Huang2014} do not obtain the well-known
  results for low-frequency (given by the horizon area) and high-frequency
  (given by the capture cross section) absorption cross section limits.
  Moreover, their results for small black hole charge do not agree with the
  Schwarzschild limit, as it should also be the case.}%
  \bibAnnoteFile{Stop}{comHuang}%
\bibitem{Huang2014}%
  \BibitemOpen
  \bibfield{author}{%
  \bibinfo {author} {\bibfnamefont{H.}~\bibnamefont{Huang}}, \bibinfo {author}
  {\bibfnamefont{P.}~\bibnamefont{Liao}}, \bibinfo {author}
  {\bibfnamefont{J.}~\bibnamefont{Chen}},\ and\ \bibinfo {author}
  {\bibfnamefont{Y.}~\bibnamefont{Wang}},\ }%
  \bibfield{journal}{%
  \bibinfo {journal} {J. Gravity}\ }%
  \textbf{\bibinfo {volume} {2014}},\ \bibinfo {pages} {231727} (\bibinfo
  {year} {2014})%
  \bibAnnoteFile{NoStop}{Huang2014}%
\end{thebibliography}%
\end{document}